\documentstyle[floats,aps,epsf,prl]{revtex}
\begin{document}
\draft
\preprint{March 27, 1999}

\twocolumn[\hsize\textwidth\columnwidth\hsize\csname
@twocolumnfalse\endcsname

\title{A new constant-pressure molecular dynamics method for finite system}
\author{D. Y. Sun}
\address
{Institute of Solid State Physics, Academia Sinica, 230031-Hefei, 
P. R. China}
\author {X. G. Gong}
\address
{Institute of Solid State Physics, Academia Sinica, 230031-Hefei, 
P. R. China \\
Department of Physics, Fudan University, Shanghai 200433, P.R. China}

\date{Received \today}
\maketitle
\begin{abstract} 
In this letter, by writing the volume as a function of coordinates of atoms, 
we present a new constant-pressure molecular dynamics 
method with parameters free.
This method is specially appropriate for the
finite system in which the periodic boundary condition does not exist. 
Simulations on the carbon nanotube and the Ni nanoparticle clearly
demonstrate the validity of the method.
By using this method, one can easily obtain the equation of states
for the finite system under the external pressure.

\end{abstract} 

\vskip1pc]

The molecular dynamics (MD) simulation method is a powerful tool, 
widely used in chemistry, physics, and materials science.\cite{AlT}
As a very important achievement, constant-pressure MD
proposed by Andersen,\cite{An}
and subsequently extended by Parrinello and Rahman,\cite{PaR} has provided a powerful 
method  to study systems under the external 
pressure. Now the constant-pressure 
MD has become  a standard tool to   
 study the physical properties  under the external pressure.
 Especially, it has played a central role in investigating the structural phase transition
as well as in geophysical and astrophysical applications.

Recently, the study of low-dimensional and biological system 
under the external pressure become a considerable interest.\cite{AnP,Ga} 
In particular,
studies on the carbon nanotube,\cite{ShJ}
clusters and nanocrystal, such as CdSe, CdS and Si 
nanocrsystals,\cite{ToA} have revealed a wealth of interesting new phenomena.
Usually computer simulation could 
substantially complement the experimental information.
However, the traditional constant-pressure MD method
was designed for the infinite system with a periodic boundary condition, 
it can not be directly applied to the  
 finite system, 
such as nanotubes and nanocrystals, 
in which the periodic boundary condition does not exist.
For this reason it is necessary to develop a new computational scheme to study
the finite system. 
Very recently, 
Martonak, Molteni and Parrinello have successfully studied the 
pressure induced amorphorpization of $Si_{35}H_{36}$ cluster 
by introducing  pressure-transmitting liquid.\cite{MaM}
In order to have a well-defined isotropic constant 
pressure on the cluster, the number of particles and the volume of the 
pressure-transmitting liquid 
should be much larger than that of the cluster, 
thus it costs
a significant overhead of the computation. Additionally, 
one also must 
determine how the liquid atoms interacts with themselves and with 
cluster atoms so that 
the liquid does not 
crystallize,  or, in the  time scale of the simulation 
the liquid does not undergo a glass transition. 

In this paper, by writing the volume as a function of coordinates of atoms,
we propose a  new constant-pressure MD, which should be appropriate for  finite
systems with  arbitrary shape.
This new constant-pressure MD is parameter free, and can be used in any system
with arbitrary shape, 
especially in nanocrystals.  
We will demonstrate its validity by classical simulations of  a carbon 
nanotube,
and small Ni particles, its extension to  $ab-initio$ 
molecular dynamics method is straight forward. \cite{SuG3}

We write the Lagrangian $L$ of a $N$-atom system as:
\begin{equation}
L=\sum_{i}^{N}\frac{{\bf p}_{i}^{2}}{2m_{i}}-(\phi(\{{\bf r_{i}}\})+P_{ext}V)
\end{equation}
where $\bf{r}_{i}$, $m_{i}$, ${\bf p}_{i}$ are the coordinates, 
mass, momentum of $ith$ atom respectively,
and $\phi$ is potential of system.
$V$ is the volume of the system, 
and $P_{ext}$ is the external pressure.

If the system obeys the Newton's mechanism, the equations of motion for
${\bf r}_{i}$ derived from the Lagrangian $L$ read,

\begin{equation}
\frac{d}{dt}(\frac{\partial L}{\partial {\bf {\dot r}}_{i}}) = \frac{\partial L}{\partial {\bf r}_{i}}
\end{equation}

Obviously the enthalpy
will be conserved.
The equations of 
motion derived from Eq. 2 produce
the constant pressure ensemble for the system, as we show bellow.
For an equilibrium system with the external pressure,
\begin{equation}
\frac{1}{3V}(\sum_{i}^{N}m_{i}v_{i}^{2}-\sum_{i}^{N}{\bf r}_{i} \cdot {\bf \bigtriangledown}\phi
-\sum_{i}^{N}{\bf r}_{i}\cdot P_{ext}{\bf \bigtriangledown}V)=0
\end{equation}
Then we have,
\begin{equation}
\sum_{i}^{N}m_{i}v_{i}^{2}-\sum_{i}^{N}{\bf r}_{i}\cdot{\bf\bigtriangledown}\phi=
\sum_{i}^{N}{\bf r}_{i}\cdot P_{ext}{\bf\bigtriangledown}V
\end{equation}

It is well known that, in statistical physics,\cite{LaL} the
volume is an additive quantities, which can be written as a summation of
the volume of individual atoms, 

\begin{equation}
V=\sum_{i}^{N}v_{i},
\end{equation}
where $v_{i}$ is the volume of the $i$th atom, which can be generally written as
a cubic homogeneous function of its nearest neighbor distance
 $r_{ij}$, i.e.

\begin{equation}
v_{i}= \sum_{j\neq i}f(r_{ij}^{3}).
\end{equation}
According to Euler theorem,
\begin{equation}
\sum_{i}^{N}{\bf r}_{i}\cdot{\bf \bigtriangledown}V=3V
\end{equation}
So from Eq. 4 we have  

\begin{equation}
P_{ext}=P_{int}=\frac{1}{3V}(\sum_{i}^{N}m_{i}v_{i}^{2}-\sum_{i}^{N} {\bf r}_{i}\cdot{\bf \bigtriangledown}\phi)
\end{equation}
where $P_{int}$ refers to the internal pressure, since  
the external pressure $P_{ext}$ is
a constant, 
$P_{int}$ is also a constant.
Thus,  by writing the volume as a function of atomic coordinates, 
 a new
constant-pressure MD is presented, in which no extra parameter is introduced.

The present  constant-pressure MD method has several advantages. First,
 this method can   
make the calculation more realistic, without needing to choose the 
$mass$ for the volume as in traditional constant-pressure MD,\cite{An}
which directly affects the time scale of the 
relaxation.
Secondly, in the present  method, the 
responding of system to the external pressure is more physical. 
This is specially important 
for the inhomgenous system. However,
 in the traditional constant-pressure MD,  
the responding of system to the external pressure 
is essentially linear, $i.e.$, 
the volume of the system is linearly scaled according to 
the difference of the internal and external pressure. 

The key to the success of the present constant-pressure MD method is to 
define the volume as a function of 
atomic coordinates. 
In the traditional constant-pressure MD method, the volume is a generalized 
coordinates, which has   equal importance as an atomic  coordinate, 
the constant pressure
 is dynamically achieved  by directly changing the volume of system.
However, in the present scheme, the  constant pressure  is  obtained
by dynamically changing the motion of each atom.

To show how well this new constant-pressure MD works   in the  
real application,  we have simulated  
 carbon nanotubes and Ni nanocrystals
under the external pressure. 

In the real simulation, to get an exact formalism for the atomic
volume is physically difficult. However, there are a varity of
sufficient approximations.
 One of the simplest and direct way is based
on the Wigner-Seitz premitive cell, we used the scaled volume of 
the atomic sphere to approximate the Wigner-Seitz premitive 
cell, which has the following
form,
\begin{equation}
v_{i}= \gamma_{i} \frac{4\pi}{3}\sum_{j\neq i}(\frac{r_{ij}}{2})^{3},
\end{equation}
where the summation runs over all the  first nearest neighbors of the $i$th atom,
$\gamma_{i}$ is a scale factor which is closely related the number of 
the nearest neighbours of the $i$th atom.  
At the high temperature or in the system where the
first nearest neighbors frequently change, we used a more sophisticated
form.\cite{SuG3}

The definition of the volume is not unique, our definition for Ni nanocrystals
is one of them. If it is necessary, one can make more elegant one.  
In the study of carbon nanotube, we fix the periodic length in axial direction.
The slight different definition of the volume for nanotubes will be presented
else.\cite{SuG2}

The interaction between carbon atoms is described
by the parametrized potential developed by Brenner\cite{Br} 
according to the Tersoff bonding formalism\cite{Te}, 
which is widely used to study mechanical properties of carbon nanotube
\cite{MaN}.
The Sutton-Chen potential is used to describe the interaction between 
Ni atoms,\cite{SuC} which is also widely used in the literature\cite{XiS}.
We find that, the obtained results
for the carbon nanotube 
and Ni nanocrystal are in good
agreement with what obtained by other methods.

The evolution of the instantaneous pressure 
in the simulation  of the (10$\times$10) carbon nanotube  and 
the Ni nanocrystal (Fig. 1), 
shows that the present method does recover a constant pressure simulation.
Although  the  instantaneous pressure  of the system 
fluctuates, their average values are equal to the set external pressure,
0.7 GPa for the carbon nanotube and 7.0 GPa for the Ni nanocrystal.
The correlation between the volume and pressure can also be clearly observed.

What shown in Fig. 2 is the calculated energy via volume (pressure) 
for carbon nanotube, the close agreement 
 between the static calculation and the present constant-pressure MD simulation
clearly demonstrates the validity of the present method.
In the static calculation,  energies are calculated with
linear scaling the
radius of the carbon nanotube without relaxing the atomic positions.
All the energies shown in the lower panel of Fig. 2 are relative to
 the minimum energies, and the volume are renormalized by the equilibrium volume
without the external pressure. 
We performed  the constant-pressure MD simulation at the zero temperature and
300 K at various external pressure from 0 GPa to 2 GPa. 
  Both the static
calculation and 
the constant-pressure MD simulation at the zero temperature
give the very similar results,
the slight  difference comes from the relaxation of the atomic coordinates.

We show
the enthalpy as a function of the reduced volume for static and present MD
results in the middle panel
 of Fig. 2. It can be seen that, 
at each volume, the 
enthalpy of present MD, in which the structures under the external pressure are
relaxed, is always smaller than that from the static
calculation as it should be.
The difference of the calculated enthalpy between the static and MD at 
0 K increasing with reducing
volume suggests the atomic relaxation becomes more and more important.
The reasonable agreement between the {\it ab-initio} and the model potential
calculation shows that the present model potential can qualitatively describe
the behavior of the carbon nanotube.
The new constant-pressure MD also 
correctly describe the finite temperature properties.
 
The  equation of states (EOS) at finite temperature 
for Ni nanocrystal can be obtained through the simulation,
 which  is  reasonable 
in agreement with the bulk phase EOS calculated 
by traditional constant-pressure MD.
Fig. 3 shows the equation of states for a  Ni nanocrystal
and bulk phase calculated
by the new constant-pressure MD and traditional MD respectively.
All the energies shown in Fig. 3 are relative to
 the minimum energies, and the volume is renormalized by 
 the equilibrium volume without the external pressure at 300 K. 
The nanocrystal and bulk phase show similar behavior.
However, from the figure, one can see that the nanocrystal is not as hard as
the bulk phase. If assuming  
 first-order Birch-Murnaghan EOS.\cite{BiM} is also appllicable to
the finite system, we yields the
 bulk modulus 136 GPa and 166 GPa for nanocrystal and bulk phase respectively,
 in agreement with our previous results.\cite{SuG1} 

In summary, by writing the volume
as a function of coordinates of atoms, 
we have proposed a new constant-pressure molecular dynamics method 
for the finite system with parameters free, where 
the external pressure  
could be exactly implemented. 
Simulations on the carbon nanotube and Ni nanoparticles have clearly
 demonstrated the validity
of the method, in which the constant pressure is recovered.
We have also shown that, with the newly proposed scheme,  
 the EOS for the finite system can be calculated through a
 molecular dynamics simulation. The extension to $ab-initio$ molecular
 dynamics method is straight forward and is in progress.

We thanks Dr. D. J. Shu and 
Dr. G. Chen
for the technical assistance.
This work is
 supported by
the National Science Foundation of China,
 the special funds for major state basic research  and CAS projects.

\begin{figure}
\centerline{\vbox{\epsfxsize=100mm \epsfbox {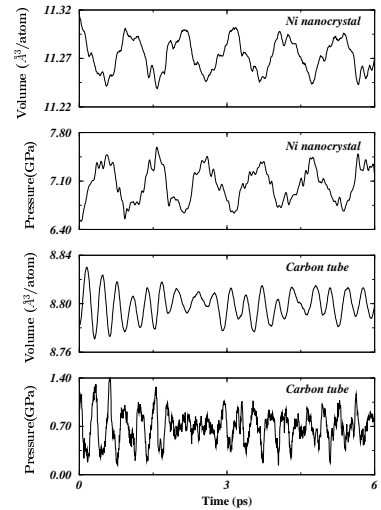}}}
\caption{The evolution of the instantaneous pressure and volume through MD
 runs for the carbon nanotube (lower two) and Ni nanocrystal (upper two).
The pressure and volume of the system
fluctuates around the average value.
The new constant-pressure MD does recover
a constant pressure.}
\label{fig2}
\end{figure}

\begin{figure}
\centerline{\vbox{\epsfxsize=100mm \epsfbox {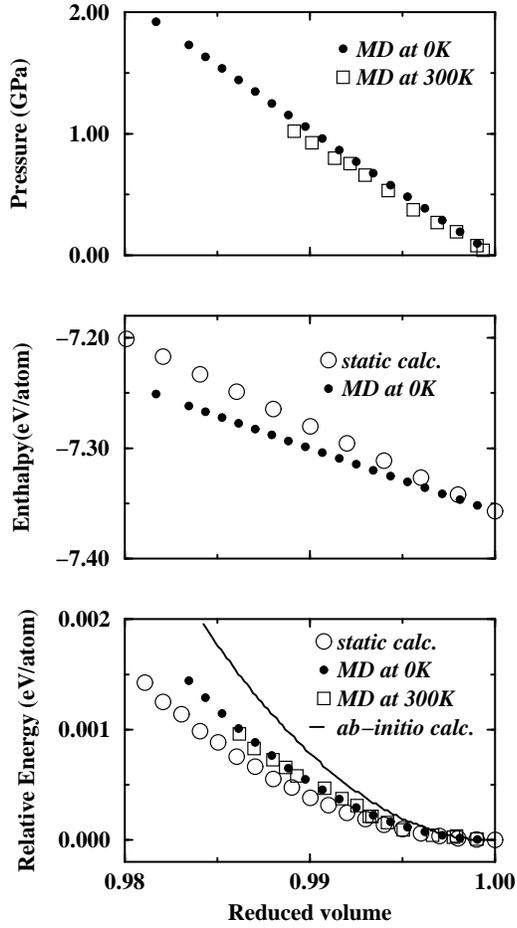}}}
\caption{The calculated properties for the carbon nanotube.
Low panel:
The energy as a function of the volume by the new
constant-pressure MD at 0 K (filled circle), 300 K (open square),
static calculation (open circle) and $ab-initio$ result (solid line).
Middle panel: The enthalpy as a function of the reduced volume 
for static (open circle)
 and present MD (filled circle) calculation.
Up panel: The pressure-volume relation of the carbon nanotube at 0 K 
(filled circle) and 300 K (open square).
Our MD results at zero temperature are in good agreement with
the static results except for large volume charge, where the
structural relaxation includes in our MD runs.
}
\label{fig1}
\end{figure}

\begin{figure}
\centerline{\vbox{\epsfxsize=100mm \epsfbox {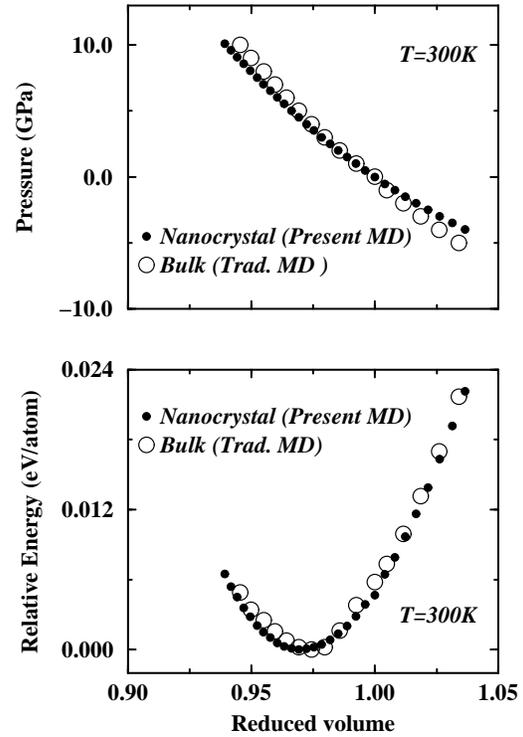}}}
\caption{
The equation of states for Ni nanocrystals (filled circles) 
and bulk phase (open circles), 
where the data for 
$Ni_{3151}$ nanocrystals and bulk phase are calculated by the new and 
traditional molecular dynamics simulation respectively. 
}
\label{fig3}
\end{figure}


\begin{references}
\bibitem{AlT} M. P. Allen and D. J. Tildesley, {\it Computer Simulation of Liquid}, Clarendon Press, Oxford, (1997).
\bibitem{An} H. C. Andersen, J. Chem. Phys. {\bf 72}, 2384 (1980).
\bibitem{PaR}M. Parrinello and A. Rahman, J. Appl. Phys. {\bf 52}, 7182 (1981);$ibid$ Phys. Rev. Lett. {\bf 45}, 1196 (1980).
\bibitem{AnP}{\it Proceedings of the XXXVI European High-Pressure Research Group Meeting on molecular and Low Dimensional System under pressure, Catania, Italy}, 1998, edited by G. G. N. Angilella, R. Pucci, and G. Piccitto, and F. Siringo, Book of Abstracts.
\bibitem{Ga}F. Gradrat {\it et al.}, Eur. J. Biochem, {\bf 262}, 900 (1999).
\bibitem{ShJ}See for example, 
Weidian Shen and Bin Jiang, Bao Shan Han and Si-shen Xie, Phys.
Rev. Lett. {\bf 84}, 3634 (2000); S. Reich, H. Jantoljak and C. homsen, 
Phys. Rev. B {\bf 61}, R13389 (2000); U. D. Venkateswran {\it et al.},
Phys. Rev. B {\bf 59}, 10928 (1999); M. J. Peters, L. E. McNeil, J. P. Lu and
D. Kahn, Phys. Rev. B {\bf 61}, 5939 (2000).
\bibitem{ToA}S. H. Tolbert and A. P. Alivisatos, Z. Phys. D {\bf 26}, 56 (1993); $ibid.$ J. Chem. Phys. {\bf 102}, 4642 (1995);  $ibid.$ Science  {\bf 265}, 373 (1994); $ibid.$ Annu. Rev. Phys. Chem. {\bf 46}, 595 (1995); S. H. Tolbert {\it et al.}, Phys. Rev. Lett. {\bf 76}, 4384 (1996).
\bibitem{MaM}R. Martonak, C. Molteni and M. Parrinello, Phys. Rev. Lett. {\bf 84}, 682 (2000).
\bibitem{SuG3}D. Y. Sun and X. G. Gong, to be published.
\bibitem{LaL} {\it Statistical Physics} 3rd edition Part 1, L. D. Landau and 
E. M. Lifshitz, Pergamon Press, 1976.
\bibitem{SuG2}D. Y. Sun, D. J. Shu and X. G. Gong, to be published.
\bibitem{Br} D. W. Brenner, Phys. Rev. B {\bf 42}, 9458 (1990).
\bibitem{Te} J. Tersoff, Phys. Rev. Lett. {\bf 61}, 2879 (1988).
\bibitem{MaN} see for example, M. B. Nardelli, B. I. Yakobson and J. Bernholc, 
Phys. Rev. Lett. {\bf 81}, 4656 (1998); 
B. I. Yakobson, C. J. Brabec and J. Bernholc, 
Phys. Rev. Lett. {\bf 76}, 2511 (1996); Yueyuan Xia, Yuchen Ma, Yuelin Xiang, 
Yuguang Mu, Chunyu Tan and Liangmo Mei, Phys. Rev. B {\bf 61}, 11088 (2000).
\bibitem{SuC}A. P. Sutton and J. Chen, Phil. Mag. Lett. {\bf 61}, 139 (1990).
\bibitem{XiS}see for example, Y. Xiang, D. Y. Sun and X. G. Gong, J. Phys. Chem.
A {\bf 104}, 2746 (2000); J. Uppenbrink and D. J. Wales, J. Chem. Phys. {\bf 96}
, 8520 (1992); A. Kara and T. S. Rahman, Phys. Rev. Lett. {\bf 81}, 1453 (1998).
\bibitem{ChG}G. Chen {\it et al.}, unpublised.
\bibitem{BiM}F. Birch, J. Geophys. Res. {\bf 57}, 227 (1952).
\bibitem{SuG1}D. Y. Sun, X. G. Gong and X. Q. Wang, submitted; D. Y. Sun and X. G. Gong, submitted.

\end{references}
\end{document}